\documentclass[epjCONF,columns]{svjour} 
\usepackage{graphics}
\usepackage{graphicx}
\usepackage[varg]{txfonts} % Times fonts
\usepackage[latin1]{inputenc}
\usepackage{amssymb}
\usepackage{epstopdf}
\session-title{The 2011 Hadron Collider Physics symposium (HCP-2011)}

\begin{document}
\title{\centering Measurement of the charge asymmetry in top quark pair production in pp collisions at $\sqrt s = 7 $ TeV using the ATLAS detector}
\author{
\centering
Rachik Soualah \thanks{\email{rsoualah@cern.ch}}}

\institute{\centering
On behalf of the ATLAS collaboration\\
INFN Gruppo Collegato di Udine and ICTP Trieste, Italy.
}

\abstract{ We present a measurement of the charge asymmetry in top-antitop production using data corresponding to an integrated  luminosity of $0.70 fb^{-1}$ of proton-proton collisions at $\sqrt s = 7$ TeV  collected by the ATLAS detector. The top pair events decaying semileptonically (lepton+jets channel) to either an electron or muon, missing transverse energy and at least four jets are selected. The reconstruction of the $t\bar{t}$ events was performed using a kinematic likelihood approach. The difference of absolute values of top and antitop rapidities is used to define the charge asymmetry: $A_{C} = (N(|\Delta Y|>0) - N(|\Delta Y|<0)) / (N(|\Delta Y|>0) + N(|\Delta Y|<0))$. To allow comparisons with theory calculations, a Bayesian unfolding technique is applied to correct the measured $|\Delta Y|$ distributions for acceptance and detector effects. The top charge asymmetry in both channels (e and mu) after correction is measured to be: $A_{C} = -0.009 \pm 0.023 (stat) \pm 0.032 (syst)$ (e+jets channel) and $A_{C} = -0.028 \pm 0.019 (stat) \pm 0.022 (syst)$ ($\mu$+jets channel) giving a combined result of : $A_{C} = -0.024 \pm 0.016 (stat) \pm 0.023 (syst)$. These results are compatible with the Standard Model predictions of $A_{C}=0.006$.
} %end of abstract
\maketitle
%

% ===========================
\section{Introduction}
\label{intro}
The top quark is the most massive elementary particle observed so far. With its mass close to the electroweak symmetry scale, it can play a special role in the Standard Model (SM) and also in many physics beyond the Standard Model (BSM) theories. The $t\bar{t}$ pair production at the hadron colliders represents one of the most probing tests of Quantum ChromoDynamics (QCD) at high energies.\\
At leading order in perturbative QCD, $t\bar{t}$ production is predicted to be symmetric under charge conjugation. At next-to-leading order (NLO), the processes $q\bar{q}\to t\bar{t}g$ and $qg\to t\bar{t}q$ exhibit a small asymmetry, due to interference between initial and final state gluon emission. The $q\bar{q}\to t\bar{t}$ process also possesses an asymmetry due to the interference between the Born and box diagrams.  It is predicted that the top quark will be emitted preferentially in the direction of the incoming quark and the antitop in the direction of the antiquark ~\cite{Ref1}. In $p\bar{p}$ collisions, the charge asymmetry can be interpreted as a forward-backward asymmetry.
For $m_{t\bar{t}}> 450,\rm{GeV}/c^2$, the CDF experiment measures an asymmetry in the $t\bar{t}$ rest frame which is 3.4~$\sigma$ above the  SM prediction~\cite{Ref2}.

At the LHC with $7$ TeV $pp$ collisions, the dominant production mechanism for $t\bar{t}$ production is the gluon-gluon fusion process which is symmetric, while the $t\bar{t}$ production via $q\bar{q}$ or $qg$ is small in most of the phase space. Nevertheless, QCD predicts a small excess of centrally produced antitop quarks while top quarks are produced, on average, at  higher absolute rapidities. This can be understood by the fact that $t\bar{t}$ production via $t\bar{t}$ annihilation is dominated by initial valence quarks with large momentum fractions whereas antiquarks coming from the sea have smaller momentum fractions. With top quarks preferentially emitted in the direction of the initial quarks in the $t\bar{t}$ rest frame, the boost into the laboratory frame drives the top mainly in the forward or backward directions, while antitops are kept more in the  central region.

This paper presents the measurement of the charge asymmetry $A_C$ using $t\bar{t}$ events decaying semi-leptonically (lepton+jets channel) to either an electron or muon and at least four jets using data corresponding to an integrated luminosity of $0.7 fb^{-1}$~\cite{Ref3}. The charge asymmetry $A_C$ is defined as: 

\begin{equation}
A_C = \frac{N(\Delta |Y| >0) - N(\Delta |Y| <0)}{N(\Delta |Y| >0) + N(\Delta |Y| <0)},
\label{Ac}
\end{equation}

where $\Delta |Y| \equiv  |Y_{t}| - |Y_{\bar{t}}|$ represents the difference of the absolute values of top and antitop rapidities ($|Y_{t}|$ and $|Y_{\bar{t}}|$ ) and $N$ is the number of events with $\Delta |Y|$ being positive or negative. The charge of the top or antitop quark is determined by the charge of the lepton coming from the leptonically decaying top quark.

% ===========================
\section{Event selection}
\label{EvSel}
The event selection of the single lepton $t\bar{t}$ final state consists of the following requirements on the reconstructed objects: single lepton (e or $\mu$) trigger, at least one primary vertex with at least 5 associated tracks, exactly one lepton ($e$ or $\mu$) with $p_{t} > 25 GeV$ for electrons and $p_{t} >20 GeV$ for muons, either matched to the trigger, $ET_{miss}>20 GeV$, $ET_{miss}+ M_{T} > 60 GeV$ (muon channel) $ET_miss>35 GeV$ and $M_{T} > 25 GeV$ (electron channel) to suppress the higher QCD multi-jet background where each event  is required to have at least 4 jets with  $p_{t}>25$ GeV  and $|\eta| < 2.5$. At least one of these jets is required to be b-tagged (SV0).

% ===========================
\section{Background determination}
\label{bck}
The estimation of the background is done by using a data-driven  technique. The method used for evaluating the QCD multijet background with fake and non-prompt leptons in both the electron and muon channels is the so called ' Matrix Method', which relies on defining loose and tight event samples and measuring the efficiencies of real ($\epsilon_{real}$)
and fake ($\epsilon_{fake}$) loose leptons to be selected as tight leptons. These efficiencies are defined as:
\begin{equation} 
\epsilon_\mathrm{real} = \frac{N^\mathrm{tight}_\mathrm{real}}{N^\mathrm{loose}_\mathrm{real}} \:\:\: \mathrm{and} \:\:\: \epsilon_\mathrm{fake} =
\frac{N^\mathrm{tight}_\mathrm{fake}}{N^\mathrm{loose}_\mathrm{fake}}, \label{eqn:intro-mm-real-fake} \end{equation}
where $N^\mathrm{loose}_\mathrm{real}$ and $N^\mathrm{loose}_\mathrm{fake}$ are the
numbers of events containing real and fake or non-prompt leptons, which pass the loose lepton requirements, $N^\mathrm{tight}_\mathrm{real}$ and $N^\mathrm{tight}_\mathrm{fake}$ are the number of real and fake lepton events passing the tight selection criteria. 
\\
A W charge asymmetry is expected where we get the shape from the MC. The Normalization factor determined from data based on the charge asymmetry for each jet multiplicity bin using the well known ratio $r(MC) = W^{+}/W^{-}$ from the simulation. To exploit the total W+jets rate from data, we use the formula:

\begin{equation}
\label{chargeaform}
N_{W^+} + N_{W^-} = \left( {r_{MC}+1 \over r_{MC}-1} \right) (D^+ - D^-),
\end{equation}
$D^{+}$ and  $D^{-}$ are the numbers of events in data after $t\bar{t}$ selection (The W charge is determined from the lepton charge). The number of the estimated W+jets events with at least one b-tagged jet is estimated as:

\begin{equation}
 \label{eq:taggedfrac}
 W_{\rm tagged} = W_{\rm pretag} \cdot f_{\rm tagged}
\end{equation}
where $W_{\rm pretag}$ is the total estimated number of $W$+jet
events and $f_{\rm tagged}$ is the fraction of $W$+jet events passing
the requirement of having at least one $b$-tagged jet, as computed
from Monte Carlo simulation.

In addition to the backgrounds mentioned previousely, there are small background contributions (the single top production, $Z+jets$ and diboson events) which are evaluated using MC normalized to the event (N)NLO cross sections.  
    
  %====================    
   
\section{Event yield}
\label{EY}
The final numbers of the remaining events after all the cuts are summarized in Table 1. In both electron and muon channels, the W+jets is the main background. The number of events in the electron channel is significantly lower than the muon channel due to the higher $p_{T}$ cut and the more stringent $Et_{miss}$ cut. Overall there is a good agreement between expectation and data.
  
    \begin{table}[htbp]
\begin{center}
    {\footnotesize
  \begin{tabular}{| l | r r | r r | r r | r r |}
      \hline
      Channel& $\mu$+j pretag &$\mu$+j tag & $e$+j pretag &  $e$+j tag            \\ \hline \hline
      $t\bar{t}$              & 4784   $\pm$    5    & 3247 $\pm$    4    &  3293   $\pm$    4    & 2218   $\pm$    4   \\
      \hline
      S. top              & 306    $\pm$    2    & 171  $\pm$    2    &  219    $\pm$    2    & 124    $\pm$    2  \\ 
      Z+jets                  & 632    $\pm$    7    & 43   $\pm$    2    &  535    $\pm$    7    & 35     $\pm$    1  \\
      Diboson                 & 90     $\pm$    2    & 8    $\pm$    1    &  56     $\pm$    1    & 5      $\pm$    0 \\
      W+jets                  & 5741   $\pm$    915  & 494  $\pm$    234  &  3436   $\pm$    628  & 309    $\pm$    144  \\
      QCD                     & 1103   $\pm$    552  & 227  $\pm$    227  &  665    $\pm$    332  & 84     $\pm$    84     \\ \hline\hline
      Tot.B        & 7871   $\pm$    1068 & 943  $\pm$    326  &  4910   $\pm$    711  & 557    $\pm$    167   \\ \hline
      S+ B     & 12655  $\pm$    1068 & 4189 $\pm$    326  &  8203   $\pm$    711  & 2775   $\pm$    167   \\ \hline \hline
      Obs.               & 2705     &4392  & 8193     & 2997      \\ \hline
    \end{tabular}
    }
    \end{center}
    \caption{The observed and expected number of events  from $t\bar{t}$ signal events and various background processes  for the pretag and tagged samples in both channels electron and muon. }
  \label{evtnumbers}
\end{table}

%=============================
\section{The $t\bar{t}$ topology reconstruction}
\label{topo}

The full kinematic reconstruction of $t\bar{t}$ tevents is performed via Likelihood maximization method to built the asymmetry observable $|Y_{t}|-|Y_{\bar{t}}|$.
The likelihood takes as inputs the measured energies, pseudorapidities
and azimuthal angles of four jets, the measured energy of the lepton,
and the missing transverse energy. Up to five jets with the largest
transverse momentum are used in the reconstruction procedure to
increase the probability of identifying the correct combination.

\section{Unfolding the top charge asymmetry}
\label{Unfo}

Unfolding is used to estimate the truth asymmetry, i.e., moving form the reconstructed asymmetry
to the truth asymmetry that would be measured with an ideal detector and infinite event statistics.
The basic principle relies on truth distribution ($T_{j}$), the reconstructed distribution ($S_{j}$) and the response Matrix ($R_{ij}$) 
defined as: 

\begin{equation}
S_{i} = \sum_{j}R_{ij}T_{j}.
\end{equation}
In this analysis, Bayes' theorem is applied iteratively in order to
invert the response matrix~\cite{Ref4}.  Regularization
is obtained automatically using a small number of iterations. The
response matrices were calculated using Monte Carlo events
generated with MC$@$NLO. The unfolding procedure was applied to the observed $\Delta |Y|$
distribution in data, after subtracting background contributions.

% =======================
\section{Results}
\label{Res}
\begin{figure}[htbp] %  figure placement: here, top, bottom, or page
   \centering
   \includegraphics[scale=0.27]{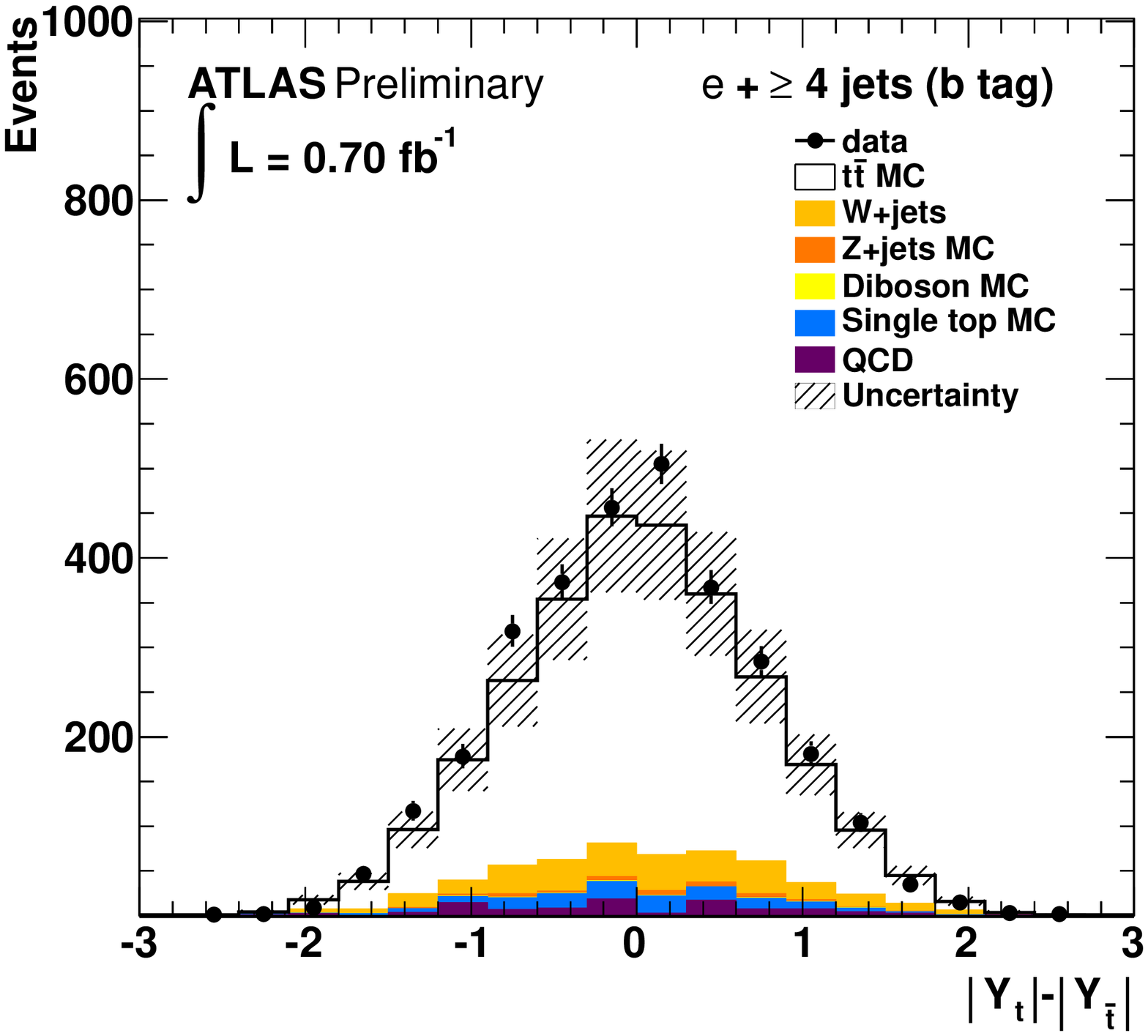}
    \includegraphics[scale=0.27]{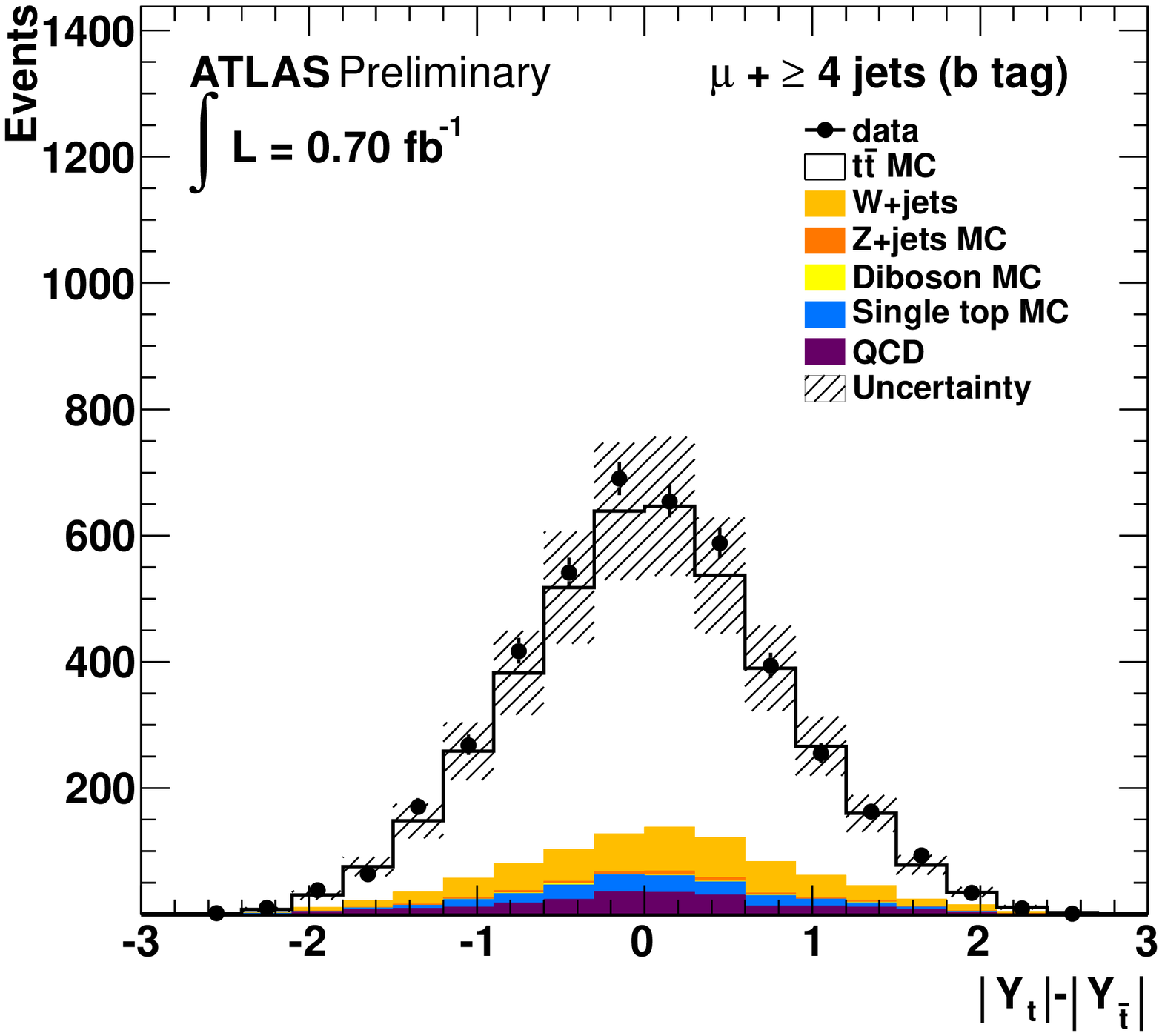} 
   \caption{The measured $\Delta |Y|$ distribution before unfolding and after b tagging is applied for the electron(up) and muon(down) channels . Data (points) and Monte Carlo estimates (solid lines) are represented. The QCD background and the normalization of the $W$+jets background are obtained using data-driven methods.}
   \label{fig:DYtop_emu}
\end{figure}

The measured distributions of the top-anti-top rapidity difference $\Delta |Y|= |Y_{t}|-|Y_{\bar{t}}|$
before the unfolding, using the event selection in Sect.~\ref{EvSel} are illustrated in Figure~\ref{fig:DYtop_emu} for both the electron and the muon channels. Figure 7 presents the corresponding  $\Delta |Y|$ after applying the unfolding.
The systematic uncertainties on the measured $A_{C}$ are listed in Table 3.
The combined results of the two channels (e: $25 \%$) and $\mu: 75\%$) using BLUE estimator give a value of:\\
\begin{equation}
A_{C} = -0.024 \pm 0.016\,\rm{(stat.)} \pm 0.023\,\rm{(syst.)}
 \nonumber
\end{equation}

\begin{table}[!htbp]
\begin{center} 
\begin{tabular}{|c|c|} 
\hline 
Asymmetry & detector and acceptance unfolded \\ 
\hline 
$A_{C}$ (e $b$-tag )& -0.009 $\pm$ 0.023 (stat.) $\pm$ 0.032 (syst.) \\ \hline
$A_{C}$($\mu$ $b$-tag) &  -0.028 $\pm$ 0.019 (stat.) $\pm$ 0.022 (syst.) \\  

\hline \end{tabular} 
\caption{\label{table:ChAsymm}The measured charge asymmetry values for the electron and muon channels before and after the $b$-tagging
requirement is applied for different levels of unfolding. The quoted uncertainties are statistical and systematic, respectively.} \end{center}
\end{table}

\begin{figure}[htbp] %  figure placement: here, top, bottom, or page
   \centering
   \includegraphics[width=2in]{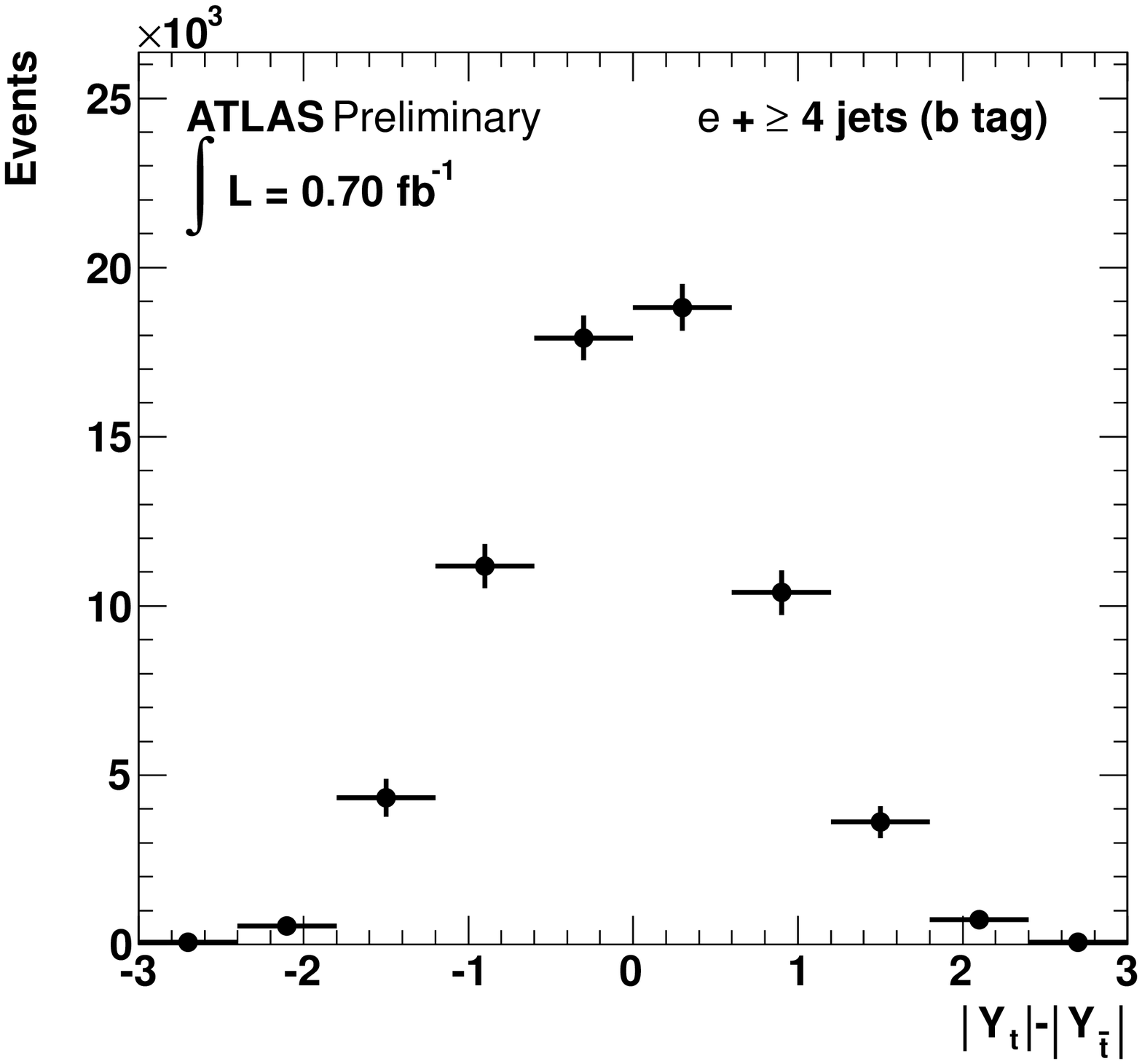} 
   \includegraphics[width=2in]{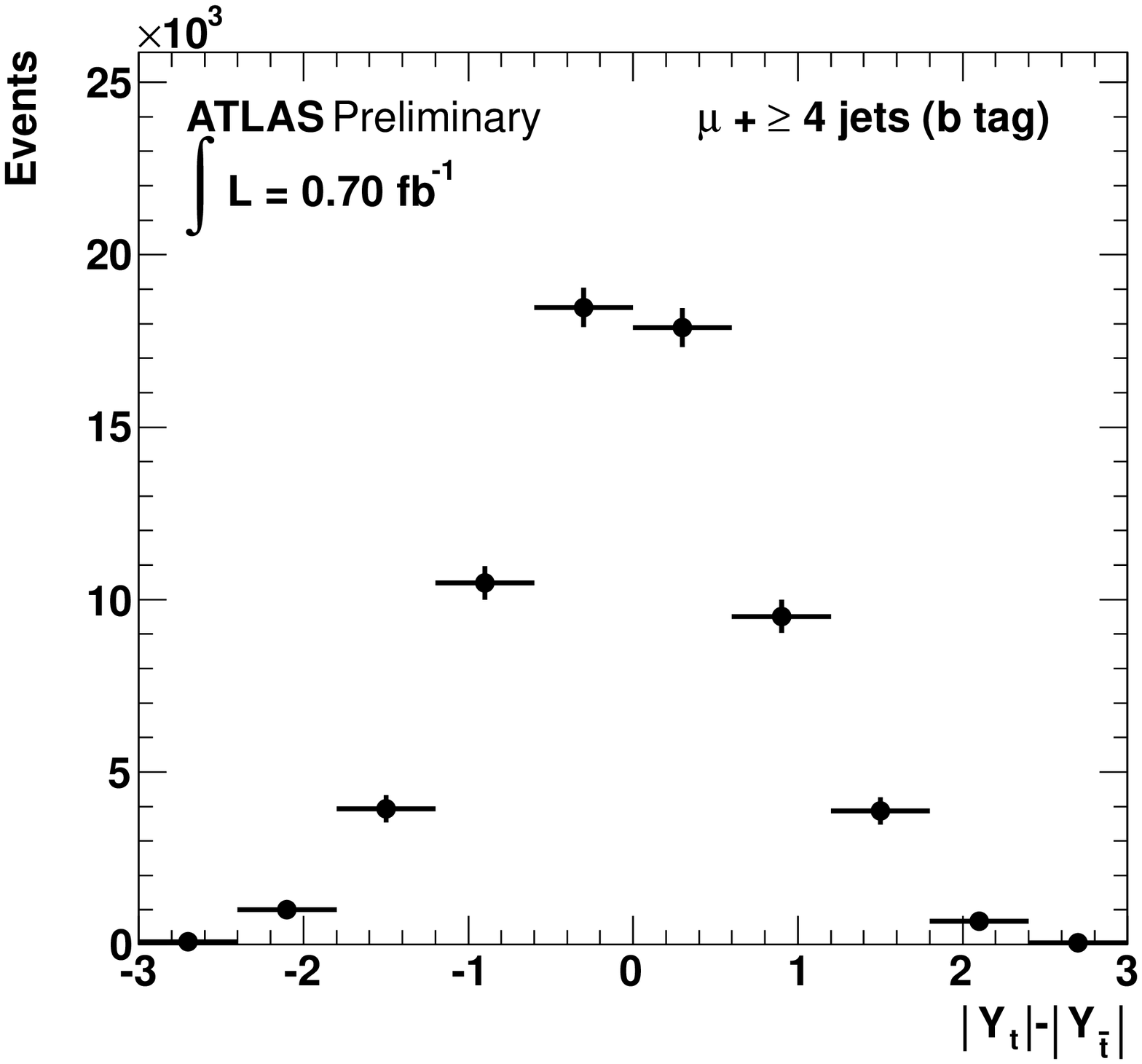} 
   \caption{The distributions of the unfolded $\Delta |Y|$ after the b-tagging for the electron (up) and for the noun (down).}
   \label{fig:example}
\end{figure}

%%===================

%===================
%\begin{figure}[htbp] %  figure placement: here, top, bottom, or page
%   \centering
%   \includegraphics[width=2in]{figures/fig_06b.eps} 
%   \caption{example caption}
%   \label{fig:example}
%\end{figure}
%===================
 
\begin{table}[htbp]
\begin{center}
\begin{tabular}{|l|r|r|}

\hline
  & $e$ channel & $\mu$ channel \\
\hline
%Source of systematic uncertainty & Electron channel & Muon channel \\
Source &  $\Delta A_C$ &$\Delta A_C$\\
\hline  
\multicolumn{3}{|l|}{\it Signal and Background modeling} \\ 
\hline
$t\bar{t}$ generator 	& 0.0243   & 0.0100 \\  
Parton Shower/fragmentation  & 0.0108   & 0.0079 \\  
ISR/FSR 	 & 0.0074   & 0.0074 \\  
PDF uncertainty  & 0.0008   & 0.0008 \\ 
Top mass 	& 0.0059   & 0.0059 \\ 
QCD normalisation   & 0.0062   & 0.0059 \\ 
W+jets normalisation & 0.0054   & 0.0097 \\
W+jets shape 	& 0.0043   & 0.0043 \\
Z+jets normalisation   & 0.0002   & 0.0002 \\ 
Z+jets shape  & 0.0010   & 0.0010 \\ 
Single Top normalisation	 & 0.0002   & 0.0002 \\ 
Diboson normalisation & 0.00001   & 0.00001 \\ 
MC sample sizes  & 0.0043   & 0.0029 \\ 
\hline  
\multicolumn{3}{|l|}{\it Detector modelling} \\ 
\hline
Muon efficiencies 			                          & (n.a.)    & 0.0002 \\  
Muon momentum scale and resolution 			          & 0.0004  & 0.0004 \\ 
Electron efficiencies 			                      & 0.0004   & (n.a.)  \\  
Electron energy scale and resolution 		 	        & 0.0004   & 0.0004 \\ 
Lepton charge misidentification 					        & 0.0002   & 0.0002 \\ 
Jet energy scale                                  & 0.0041   & 0.0046 \\  
Jet energy resolution                             & 0.0105   & 0.0040 \\ 
Jet reconstruction efficiency                     & 0.0003   & 0.0003 \\  
$b$-tagging scale factors 			                  & 0.0038   & 0.0038 \\  
Charge asymmetry in $b$-tagging efficiency        & 0.0007   & 0.0007 \\
Calorimeter readout 				                      & 0.0015   & 0.0029 \\ 
\hline
Combined uncertainty                              & 0.032\phantom{0} & 0.022\phantom{0} \\
\hline
\end{tabular}
\end{center}
\caption{ List of sources of systematic uncertainty and their impact
  on the measured asymmetry in the electron and muon channel. In cases where asymmetric 
uncertainties were obtained, a symmetrisation of the uncertainties was performed.}
\label{Table_Systematics}
\end{table}

\section{}
\label{intro}

\end{document}